\documentclass[11pt]{article}
\usepackage{bm}
\usepackage[margin=1in]{geometry}
\usepackage[utf8]{inputenc}
\usepackage{amsfonts}
\usepackage{amssymb}
\usepackage{graphicx}
\usepackage{amsmath}
\usepackage{ragged2e}
\usepackage{empheq}
\usepackage{tikz}
\usetikzlibrary{arrows,positioning,shapes,backgrounds,fit}  
\usetikzlibrary{tikzmark,calc}
\usepackage{caption}
\usepackage{float}
\usepackage[normalem]{ulem}
\usepackage{setspace} 

\date{}

\begin{document}

\title{Dynamic condensates in aggregation processes with mass injection}

\maketitle

\begin{center}
\author{\underline{Arghya Das}$^1$*\let\thefootnote\relax\footnotetext{*Corresponding author: arghya2009das@gmail.com} and Mustansir Barma$^1$\\ {\it $^1$TIFR Centre for Interdisciplinary Sciences, Tata Institute of Fundamental Research, Gopanpally, Hyderabad, 500046, India}}
\end{center}

\begin{abstract}
The Takayasu aggregation model is a paradigmatic model of aggregation with mass injection, known to exhibit a power law distribution of mass over a range which grows in time. Working in one dimension we find that the mass profile in addition shows distinctive  {\it dynamic condensates} which collectively hold a substantial portion of the mass (approximately $80\%$ when injection and diffusion rates are equal) and lead to a substantial hump in the scaled distribution. To track these, we monitor the largest mass within a growing coarsening length. An interesting outcome of extremal statistics is that the mean of the globally largest mass in a finite system grows as a power law in time, modulated by strong multiplicative logarithms in both time and system size. At very long times in a finite system, the state consists of a power-law-distributed background with a condensate whose mass increases linearly with time.
\end{abstract}
~
{\bf Keywords:} Takayasu model, coarsening, extreme values, aggregation process, nonconserved process, power laws

\pagebreak

\section{\label{sec:level1}Introduction}

The kinetics of aggregation is important in a large number of natural and experimental settings. Examples include the formation of gels in a system of branched polymers  \cite{ziff-stell}, colloidal aggregates \cite{Meakin}, aerosols \cite{Friedlander,Drossinos} and biomolecular condensates \cite{Brangwynne2011,Banani2017}. Generally speaking, there are two sorts of questions of interest. The first has to do with the geometrical structure of the aggregate itself, which is often a fractal. The second question has to do with the mass distribution of clusters, which often has interesting features such as a power law distribution or the occurrence of very large aggregates. It is the second question that is of primary interest in this paper. 

\subsection{Models of aggregation}

A widely used approach to find the distribution of aggregating masses at the mean field level is to use a master equation to account for the coalescence of clusters with masses $m$ and $m'$ \cite{F.Leyvraz,Ben-Naim-book,Rajesh-source-2004}. In this approach, a key role is played by the reaction kernel $K(m,m')$ which describes the merger of the two clusters. The kernel is proportional to the reaction rate, and its dependence on $m$ and $m'$ is chosen to embody the physics of the particular coalescence process being modelled. Amongst the different sorts of states that can result is the interesting possibility of the formation of a gel, or infinite aggregate, in a finite time \cite{ziff-stell,ziff80}.

The mean field approach discussed above neglects the spatial distribution of clusters and the fact that the two clusters must meet before they coalesce. Approaches which model spatial dispersal must also include a description of the motion of clusters; this is often taken to be diffusive. In low dimensions, diffusion can control the rate of aggregation and lead to results which differ from the mean field theory, though this is not always the case. These features are brought out in the models to be discussed below.

\begin{figure*}[h]
\begin{center}
\begin{minipage}{0.45\textwidth}
 \includegraphics[width=6.5cm, height=4.5cm]{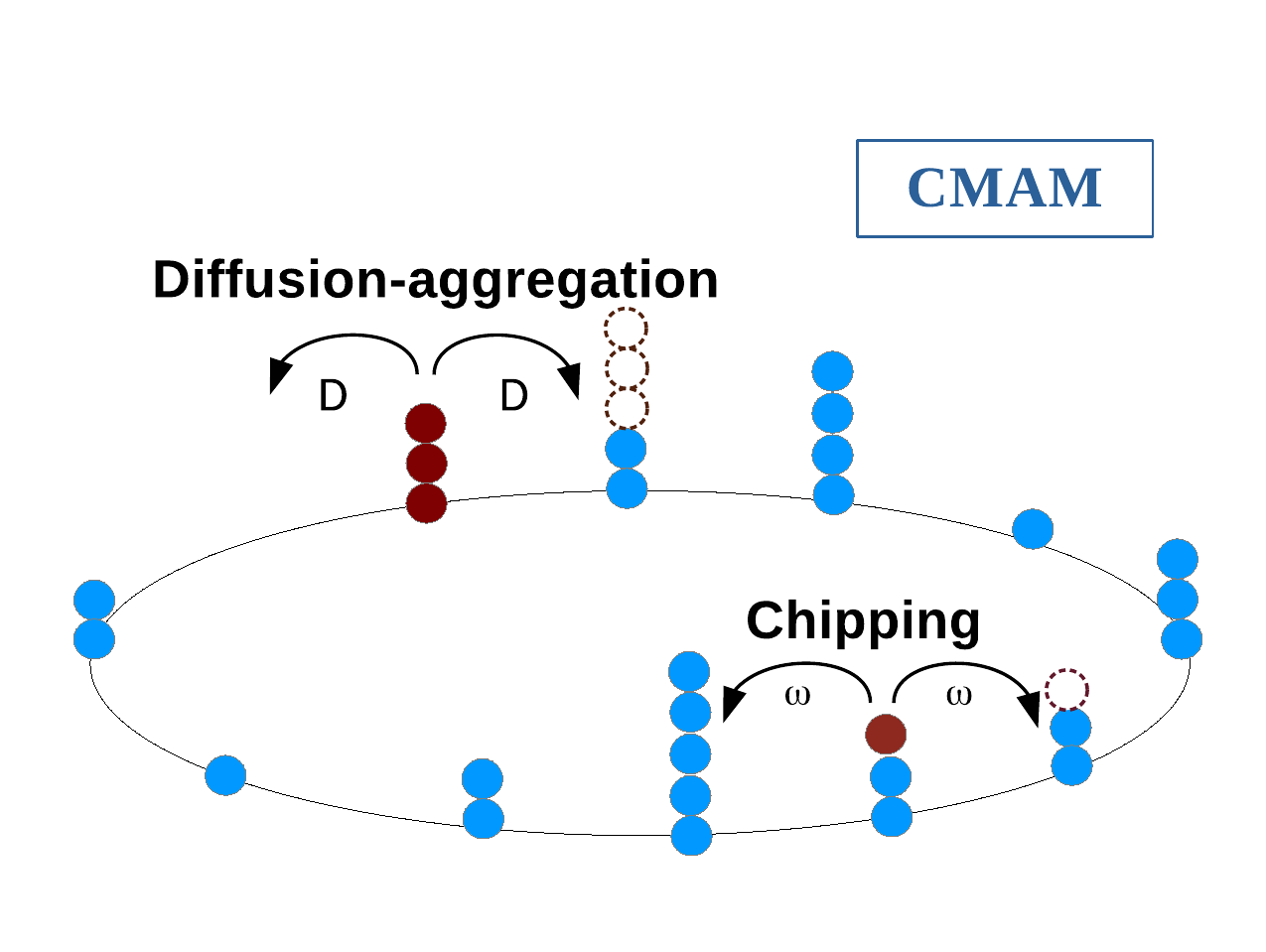}
 \caption{\it \raggedright Schematic of the kinetic moves in CMAM. The diffusion-aggregation move occurs with rate $D$ and single-particle chipping with rate $w$.} \label{cmam_schematic}
\end{minipage} \quad
\begin{minipage}{0.45\textwidth}
\includegraphics[width=6.5cm, height=4.5cm]{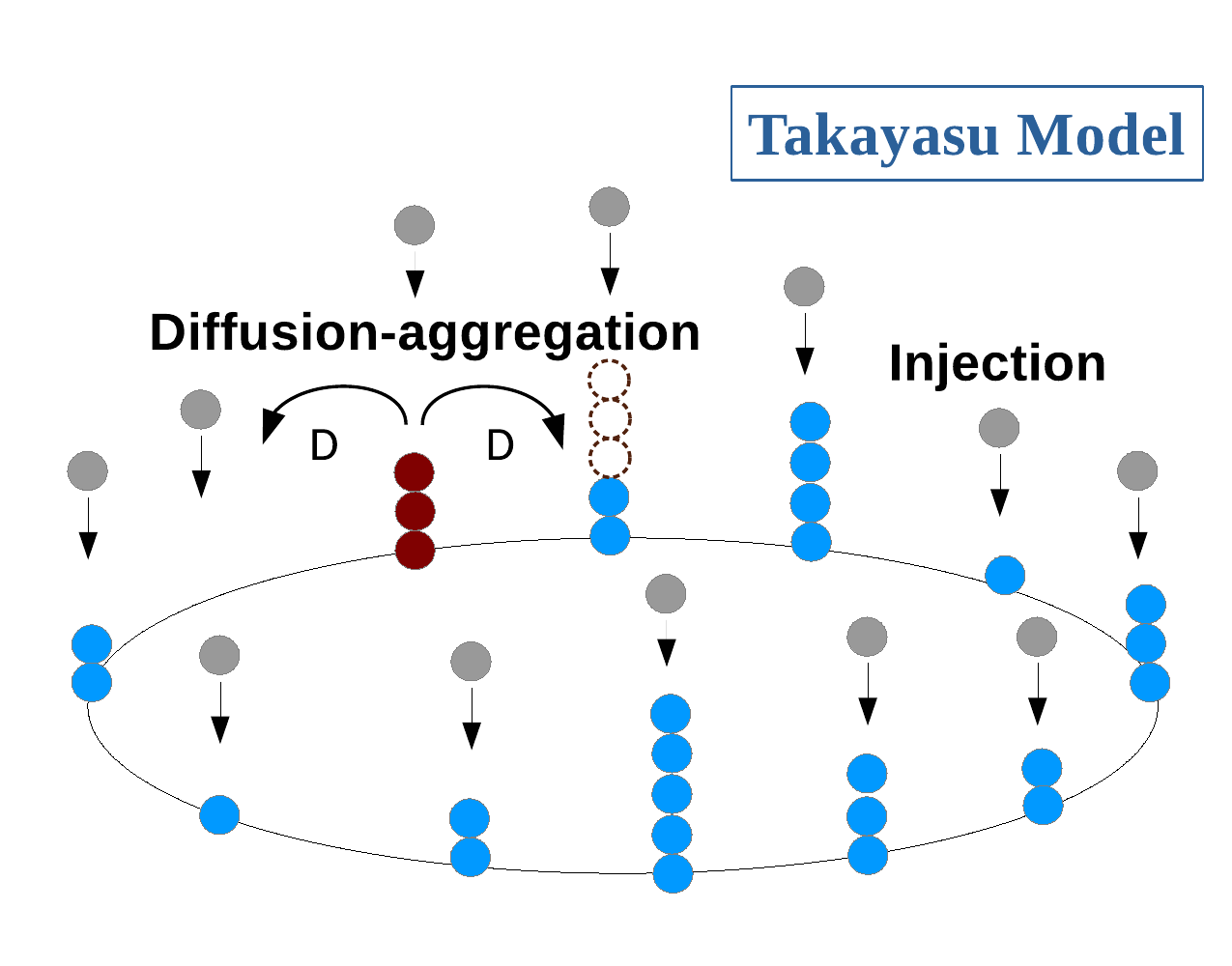}
\caption{\it \raggedright Schematic of the diffusion cum aggregation and injection moves in the symmetric version of the Takayasu aggregation model. Particle injection occurs at rate $1$.} \label{Takayasu_schematic}
\end{minipage}
\end{center}
\end{figure*}

Systems which are closed and in which mass is conserved are expected to behave quite differently from open systems which allow mass input or exchange with the environment. A model of the former variety is the conserved mass aggregation model (CMAM) in which the elementary moves include diffusion, coalescence on contact $(m, m' \rightarrow m+m')$ and chipping, or fragmentation of a cluster of size $m$ into clusters of size $m-1$ and 1 (Fig. \ref{cmam_schematic}) \cite{cmam98,Rajesh-Satya-2001}. In the steady state, this model shows an interesting phase transition from a disordered phase with exponentially distributed clusters, to a phase with a single condensate, or macroscopic cluster, in a sea of power-law distributed clusters. The phase transition, which bears a mathematical resemblance to Bose-Einstein condensation, hinges on the total mass in the system being conserved \cite{sm-jsp2000}. Other variants of the CMAM also show phase transition of the same nature \cite{Pradhan2015}. On the other hand, introducing a mass-dependent diffusion rate can curb the transition to a state with a condensate \cite{Rajesh2002}.

If the CMAM is generalized to include mass injection and ejection at the boundaries, it continues to exhibit a phase transition. However, in steady state the system shows macroscopic mass fluctuations in the condensate phase \cite{Himani-prl,Himani-jsp}. On the other hand, if mass injection occurs in the bulk of the system, quite a different physical state ensues, as discussed below. 

\subsection{The Takayasu aggregation model}

In the Takayasu model of aggregation, discussed in detail in section \ref{sec:level3}, in addition to diffusion and coalescence upon contact, there is a steady input of particles at every site of a lattice, at a constant rate \cite{takayasu88,takayasu89}. As a result, the total mass in the system increases linearly with time $t$. At the same time, the mass clusters are known to organize themselves in such a way that their mass distribution approaches a power-law form $P(m) \sim m^{- \tau}$  with $\tau=4/3$ in one dimension \cite{takayasu88,Huber91} --- an example of self-organized criticality. An exact calculation of the two-point correlation function with both random sequential and parallel dynamics \cite{Rajesh2000} shows that it follows a scaling form with the coarsening length scale ${\mathcal L(t)} \sim t^{1/z},~z=2$ \cite{Bray1994}, measured in terms of the lattice spacing.

An interesting point about the Takayasu aggregation model is that it has exact correspondences to several other models of statistical physics. Examples include the voter model \cite{Ligget}, models of river basins \cite{Scheidegger,Rinaldo}, and models of stress propagation in bead packs \cite{Coppersmith}. Thus results for the Takayasu aggregation model have wide applicability.

\subsection {Dynamic Condensates}

\begin{figure}[H]
\begin{center}
    \includegraphics[width=7.5cm, height=5cm]{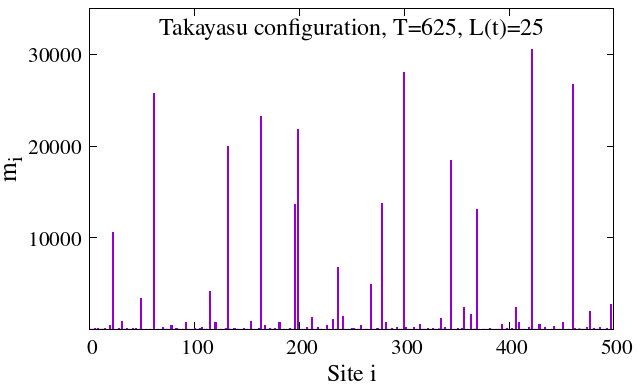}\hspace{1cm}
    \includegraphics[width=7.5cm, height=5cm]{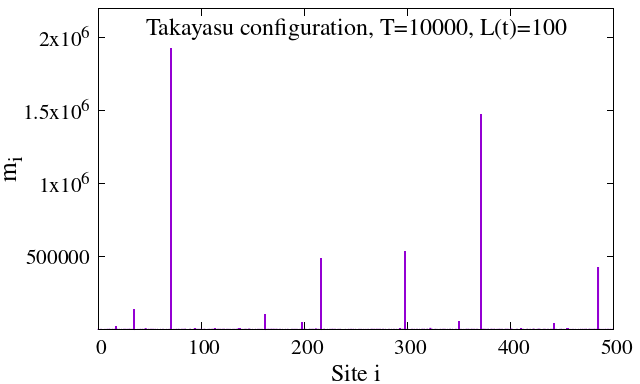}
\caption{\it \raggedright Configurations at different times in the Takayasu aggregation model with both diffusion and injection rates taken to be unity. Note the occurrence of dynamic condensates that grow in size but reduce in number in a given interval as time passes.}\label{Takayasu_configuration}
\end{center}
\end{figure}

In the context of real-space condensation \cite{Evans2005}, recall that a condensate site is defined as that site which holds a finite fraction of the total mass in steady state. In a coarsening system, we define a {\it dynamic condensate} by analogy, namely a site which holds a finite fraction of the mass in a region of size ${\cal L}(t)$. This is sensible as each region of this size is (approximately) statistically independent of other such regions. The notion of dynamic condensate is applicable to any statistical system which is coarsening towards a steady state with a condensate. As time passes and coarsening proceeds, the number of condensates falls, and the typical mass in each grows (Fig. \ref{Takayasu_configuration}).  

Extremal statistics \cite{Tippett,Arnab-SM} provides a useful way to probe the properties of large stretches and condensates, not only in steady state \cite{Frachebourg,Evans-Zia-2006,Barkai2020} but also during coarsening \cite{Iyer,cdmax}.
For instance, in a study of coarsening in the CMAM, dynamic condensates were tracked through an extremal quantity, namely the largest mass within the coarsening length ${\mathcal L(t)}$ \cite{Iyer}. The resulting distribution of the extremal mass shows interesting scaling properties, quite different from those in the steady state. Interestingly, a parallel behaviour is also found in the zero-range process, in which clusters do not diffuse as a whole, but chipping occurs at a rate which depends on the cluster mass \cite{Evans2005,Spitzer}.

In the Takayasu aggregation model, in time $t$ the total influx of mass into a region of size ${\cal L}(t) \sim t^{1/2}$ is $M(t) \sim t^{3/2}$.
Unlike the CMAM, in which the mechanism for condensate formation can be traced to mass conservation, there is no conservation law in the Takayasu aggregation model. Despite this, we find that in one dimension there is a strong tendency towards forming dynamic condensates, and they hold a finite fraction of the total mass in the system. Dynamic condensates lead to a noticeable bump at the edge of the full mass distribution (Fig. \ref{pmt_takayasu}). The mass held by the corresponding clusters amounts to a large fraction ($\sim 80 \%$ for the same diffusion and injection rate) of the whole.

\subsection{Global maximum mass}

How does the mass of the {\it globally largest} condensate grow during coarsening, in a finite system of size $L$ ? At any time $t$, we need to find the largest of the dynamic condensates in the system. Applying extremal statistics to the dynamic condensate distribution, a simple argument reveals that there is a significant {\it multiplicative logarithmic correction} to the power law growth of this quantity. Our argument follows that used in a recent study of a fluctuating interface model \cite{cdmax}. As discussed in Section \ref{sec:level5}, these logarithmic corrections have interesting implications for other models which correspond to the Takayasu aggregation model. Finally for long enough times $(t>L^z)$, the system reaches a state with a power-law distributed fluid which coexists with a moving, growing giant condensate whose mass increases linearly in time.

The paper is arranged as follows. In section \ref{sec:level2} we briefly review the steady state and coarsening in the conserved mass aggregation models. In section \ref{sec:level3}, we introduce the Takayasu aggregation model, a model of aggregation with mass injection, and discuss some of its properties. Our results for the long-time state in a finite system are given in section \ref{sec:levelss}. Next the properties of the dynamic condensates in the coarsening regime of the Takayasu aggregation model and the behaviour of the largest condensate mass are discussed in section \ref{sec:level4}. Finally, we conclude in section \ref{sec:level5}.

\section{\label{sec:level2} Conserved mass aggregation models}

In the CMAM, the system evolves through the interplay of {\it diffusion} (the entire mass of a site moves to one of its neighbouring sites and adds to the mass already present there) and {\it chipping} (a fraction of mass in a site chips off and adds to the mass of one of its randomly chosen neighbours).

 Let the rates for diffusion and chipping be $D$ and $w$ respectively (Fig. \ref{cmam_schematic}). The density is $\rho=N/L$ where $N$ is the conserved number of particles and $L$ is the system size. In the steady state, a phase transition occurs at density $\rho_c =\sqrt{1+w/D}\,\,-1$ \cite{cmam98,Rajesh-Satya-2001,sm-jsp2000}. For $\rho < \rho_c$ the mass distribution at large masses follows an exponential decay; this is the disordered or fluid phase. At the critical density the distribution follows a power law, $P(m)\sim m^{-5/2}$. For $\rho>\rho_c$, while the critical fluid remains, a condensate with average mass $(\rho - \rho_c)\,L$ forms at one site in the system, an outcome of mass conservation.
 %\footnote{Mass dependent diffusion is also investigated in [ref..] and it is found that there is no finite critical density at the thermodynamic limit.}. 
 The fluctuation properties of the condensate in CMAM have not been characterised analytically, but simulation results suggest that the distribution $P_{\rm ss}(M_c)$ of the condensate mass $M_c$ is a scaling function of $(M_c - M_0)^{\beta}$, where $M_0\approx (\rho - \rho_c)\,L$ and $\beta \simeq 0.7$ \cite{Iyer,finite-L-cmam}. This implies that all the cumulants, including the standard deviation, of the condensate mass scale as $L^{\beta}$.
 
We now give an argument \cite{finite-L-cmam} that the exact value of the exponent $\beta$ is $2/3$. Suppose $\phi$ is the cutoff exponent for the mass distribution in the {\it fluid}: $P(m) \sim m^{-\tau} f(m/L^\phi)$.
The second moment of $P(m)$, and therefore the variance of the single site mass distribution is, $\sigma^2(m) \sim L^{\phi\,(3-\tau)}$. Consequently, the variance of the mass in the fluid is $\sim L^{\phi\,(3-\tau)+1}$. Because of the conservation of total mass, the variance of the condensate mass $M_c$ also scales the same way. Now, the normalisation condition of the mass distribution induces the exponent relation, $\phi(\tau-1)=1$ \cite{Rajesh-Satya-2001}, which then leads to $2\,\beta = \phi\,(3-\tau)+1 = 2/(\tau-1) \Rightarrow \beta=(\tau-1)^{-1}$. For the CMAM, $\tau=5/2$, which then implies $\beta = (\tau-1)^{-1}=2/3$.

Recently, the coarsening to the condensate phase in CMAM was studied in \cite{Iyer} and the occurrence of dynamic condensates was established. The existence of a growing coarsening length $\mathcal{L}(t)$ is manifest in the equal time two-point correlation $G(r,t) \equiv \langle m_i(t) m_{i+r}(t) \rangle - \rho^2$ (with $\langle m_i(t)\rangle = \langle m_{i+r}(t)\rangle = \rho$) which is found to be a function of scaled separation at large separation and late times,
\begin{equation}
\lim_{r\rightarrow \infty} \lim_{t\rightarrow \infty} G(r,t) = g[r/\mathcal{L}(t)],
\end{equation}
with $\mathcal{L}(t)~\propto~t^{1/z}$ where $z=2$ in CMAM.
It is observed that while coarsening, several dynamic condensates form in the system. Their number decreases while the mass of the surviving ones increase with time. This is also corroborated by the emergence of a `bump' in the mass distribution $P(m,t)$ while coarsening (Fig. 6 of \cite{Iyer}). The standard deviation of the mass of the dynamic condensates which is measured by the largest mass in a subsystem of size $\mathcal{L}(t)$ is proportional to $\mathcal{L}(t)$ itself. The distribution $P(m^*,t)$ of the largest mass $m^*$ in a subsystem is double-peaked: the peak at smaller masses scales like $\mathcal{L}(t)^{\alpha},~\alpha\sim 0.7$ corresponding to subsystems in which there is no dynamic condensate, while the peak at larger masses scales like $\mathcal{L}(t)$ signifying the presence of dynamic condensates. This scaling further implies that not only the mean and standard deviation but all the cumulants of the dynamic condensate mass are proportional to $\mathcal{L}(t)$ in contrast to the steady state where they grow sublinearly in $L$ \cite{Iyer}. 

Similar features during coarsening are seen (see \cite{Iyer}) for another subclass of conserved mass transport models which have a high density phase with a condensate, viz. the zero range process (ZRP), with a particular dependence of the chipping rate on the mass at a site \cite{Evans2005,Evans-Zia-2006}.

\section{\label{sec:level3} Models with mass injection: Takayasu aggregation model and variants}

The Takayasu aggregation model is a paradigmatic model of nonconserved mass aggregation dynamics where mass is injected at a finite rate in the bulk of the system. This model is also connected to several other models and physical processes as discussed below. We consider the one-dimensional model defined on a lattice ring of size $L$. The kinetic moves include particle injection at a randomly chosen site with rate $1$ and diffusion of the mass from a randomly chosen site to one of its neighbours with a rate $D$.
A schematic of the model is shown in Fig. \ref{Takayasu_schematic}.

It is remarkable that although the mean mass keeps increasing due to injection, the distribution reaches a {\it statistical steady state} with a power law distribution \cite{takayasu89} as time $t\rightarrow \infty$ as a result of the competition of the the injection and aggregation dynamics. This can be understood as follows. With the diffusion-aggregation move only, at large times all the masses would coagulate and form one giant aggregate. The resulting distribution is not very interesting. To have a nontrivial distribution one needs to replenish the mass at the lower values, which can be done either by chipping as in CMAM, or by injection of particles as in the Takayasu aggregation model.

In the CMAM the competition between chipping and aggregation gives rise to an interesting distribution, and only when the relative rate of these two processes is fine tuned to the critical value, we get a pure power law behaviour. This power law in CMAM persists beyond criticality, but in addition a condensate appears at a mass proportional to the system size. In the Takayasu aggregation model, the continuous injection of particles allows coagulation at all scales, and the injection rate drives the effective aggregation dynamics at large scales leading to the power law distribution. It has also been argued that the distribution is not only statistically steady, but also stable against perturbations \cite{Privman}.

In the thermodynamic limit the $t\rightarrow \infty$ asymptotic behaviour of the single site mass distribution is exactly known, 
\begin{equation}
 \lim_{m\rightarrow \infty} \lim_{t\rightarrow \infty}\, P(m,t) \rightarrow m^{-\tau}, \label{dist_takayasu}
\end{equation}
with $\tau=4/3$.
At large but finite times, the power law behaviour is modulated by a time dependent cut-off mass which grows as $t^{\delta}$ in time, with $\delta=3/2$ \cite{Sire1993}.
We noticed an interesting feature in the distribution, namely a significant `bump' near its tail. This turns out to be due to dynamic condensates, and is discussed in detail in section \ref{sec:level4} (Fig. \ref{pmt_takayasu}).

\begin{figure}[H]
\begin{center}
    \includegraphics[width=12cm, height=8cm]{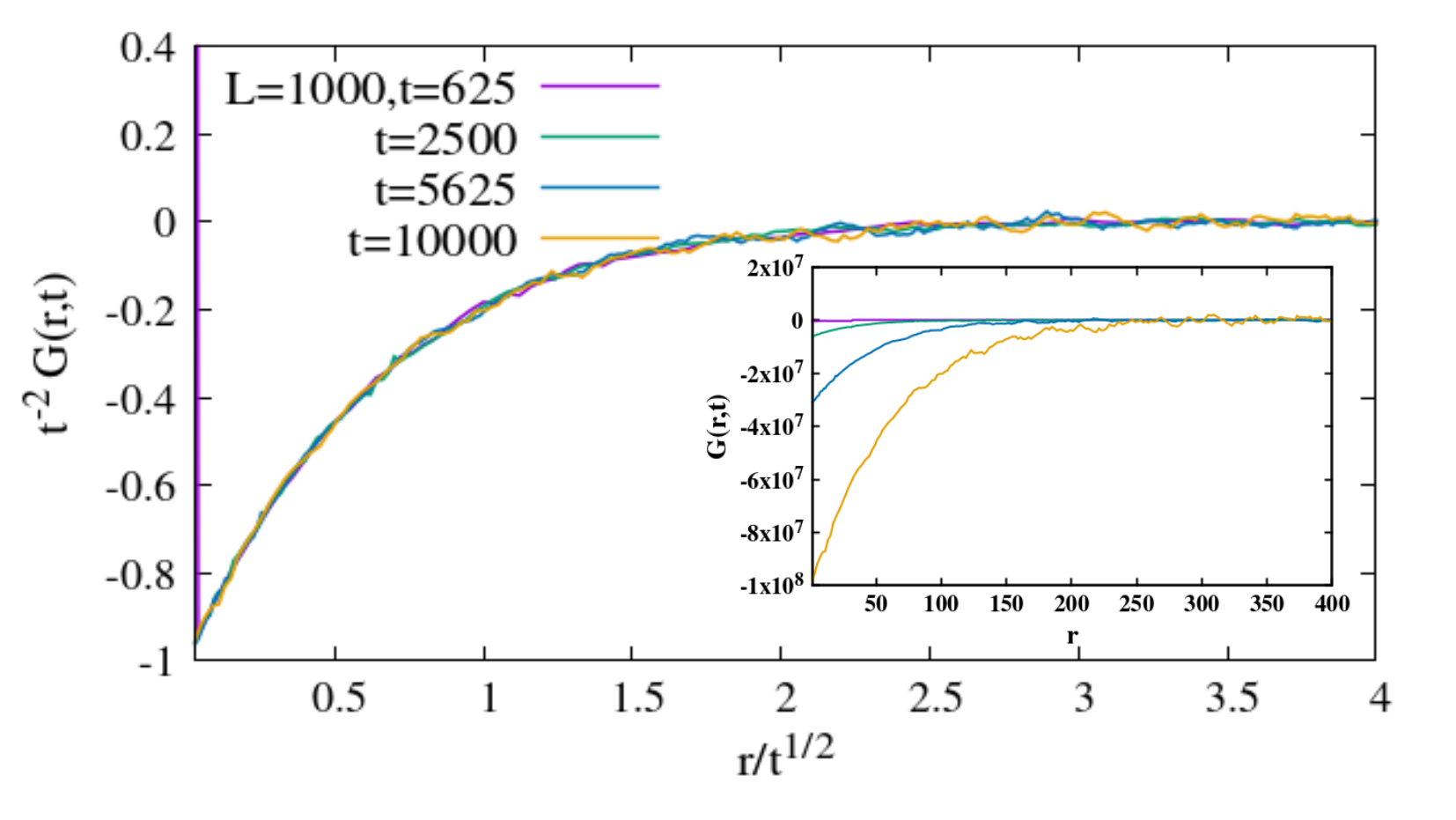}
\caption{\it \raggedright The scaled two-point correlation in the coarsening regime of the Takayasu aggregation model. The unscaled correlation is shown in the inset. Note that there is a large positive peak at $r=0$, that corresponds to the single site mass variance.}\label{corr_takayasu}
\end{center}
\end{figure}

Further, the two-point spatio-temporal correlation of the mass fluctuations in the Takayasu aggregation model has been evaluated \cite{Rajesh2000}. In particular, the equal-time two point correlation obeys a scaling form,
\begin{equation}
 G(r,t) = t^2 \,g(r/\sqrt{t}), \label{correl}
\end{equation}
in the limit $r\rightarrow \infty,t\rightarrow \infty ~{\rm with}~ r/\sqrt{t}={\rm finite}$. Here, $\rho(t)=\langle m_i(t) \rangle=t$ because of the continuous injection of mass. 
For asymmetric parallel update dynamics, the scaling function $g(u)$ has been evaluated in closed form \cite{Rajesh2000}.
With symmetric random sequential dynamics, we studied the equal time two-point correlations using Monte Carlo simulation and found similar scaling behaviour (Fig. \ref{corr_takayasu}), but the scaling function is not the same.
Figure \ref{corr_takayasu} indicates that masses with separations less than $\mathcal{L}(t)$ are strongly correlated with each other, while at separations larger than $\mathcal{L}(t)$ they are essentially uncorrelated.

Although most of the results for the Takayasu aggregation model outlined so far were obtained with asymmetric parallel update dynamics where all sites are updated simultaneously at every time step, the results are independent of the update rules \cite{Rajesh2000}. In the work presented in the next section, we have used continuous time random sequential update with symmetric dynamics.
%We have checked that, although the detailed functional form of the equal time two-point correlation is different in these two cases, the scaling behaviour of this quantity is the same. Further, the mass distribution at different times in these two cases are identical.

\subsection{Correspondences and extensions of the Takayasu aggregation model}

We now mention a few natural generalizations of the Takayasu aggregation model.

The first consists of replacing the positive quantity (mass) whose aggregation properties are considered in the Takayasu process, by a signed quantity which may be interpreted as charge. The algebraic sum of  two charges in considered in the aggregation step, while single particle injection involves either charge. The solution of this problem shows that Eq. \eqref{dist_takayasu} still holds, but the value of the exponent $\tau$ is revised from 4/3 to 5/3 when the average total charge is zero \cite{Privman,Sire1993}. With pair injection of a positive and negative charge on adjacent sites, the value of $\tau$ changes to $2$ \cite{Takayasu-pair-91,Takayasu-pair-94}.

The second generalization is the in-out model which allows not only single-particle input (deposition) as in the original Takayasu aggregation model, but also the random removal of a single particle (evaporation) at a certain rate \cite{inout2000}. If the evaporation rate is large enough, the system settles into a state with a mean particle density which does not increase in time. However, if the evaporation rate falls below a critical value, the mean particle density increases linearly in time, and $P(m)$ follows Eq. \eqref{dist_takayasu} with $\tau = 4/3$, as in the original model with no evaporation. At the critical point which separates these two phases, numerical simulations and analytical results show that Eq. \eqref{dist_takayasu} continues to hold but with $\tau = 11/6$ in one dimension \cite{Rajesh2004}. The transition point can be found within mean-field theory \cite{inout2000} and bounds on its location have been established in arbitrary dimensions \cite{Connaughton2010,Connaughton2013}.

Part of the importance of the Takayasu aggregation model stems from the fact that it is related to several processes involving either stochastic evolution or random paths \cite{Privman}. An example of the former type is the voter model, which describes the build-up of opinion in an assembly of communicating voters \cite{Ligget}. Two examples of the latter type of process are (a) the Scheidegger model of river networks \cite{Scheidegger,Rinaldo}, and (b) stress propagation in randomly packed assemblies of beads \cite{Coppersmith}. In both these cases, a typical space-time history of the Takayasu aggregation model is re-interpreted as a collection of merging paths in two dimensions, with time in the Takayasu aggregation model interpreted as a spatial coordinate. 

In (a), the collection of merging paths describes the geography of a river basin, with Takayasu world lines representing tributaries \cite{Privman}. The aggregation step describes the merger of tributaries, while particle injection captures the effect of rainfall in the river context. The mass gathered is then a measure of the area $A$ of the basin, implying a power law  ($\sim A^{-4/3}$) distribution of basin areas \cite{Scheidegger,Privman}. Further, the set of paths along which the river flows constitutes a spanning tree, which in turn maps onto the directed Abelian sandpile model \cite{Deepak}.

(b) The second example concerns the pattern of stress transmission in a vertical stack of packed beads. Each bead in a horizontal layer transmits a random fraction of stress to each of two touching beads below it. The weight supported by any bead is thus the sum of two inputs from its neighbours above (corresponding to aggregation), and its own weight adds on to the stress it transmits to the beads in the layer below it (corresponding to injection). This correspondence helps to explain the experimentally measured inhomogeneous distribution of stresses in the bottommost layer \cite{Coppersmith}.

Another interesting perspective is provided by viewing cluster-cluster aggregation as leading to a cascade towards increasing mass, analogous to the energy cascade in turbulence \cite{cascade1,cascade2}. The input of single particles replenishes the pool of small clusters and sustains the cascade. In low enough dimensions, the mass distribution is found to exhibit multiscaling. Further, in more general models, a constant flux of mass determines the scaling property of correlation functions which are linked to the flux \cite{flux1,flux2,flux3}.

\section{\label{sec:levelss} Long-time state}

\begin{figure}[H]
\begin{center}
    \includegraphics[width=12cm, height=8cm]{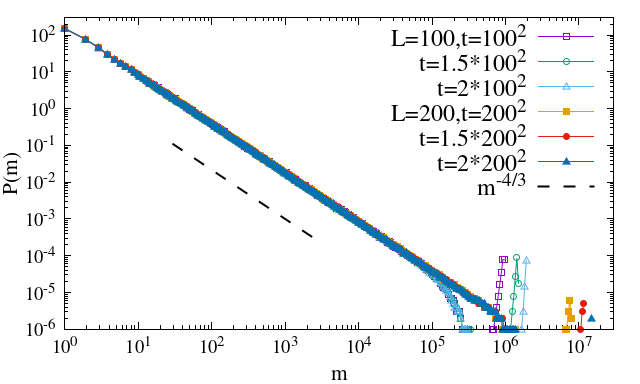}
\caption{\it \raggedright The single site mass distribution in the long-time state of the Takayasu aggregation model, for system sizes $L=100,200$. For both system sizes, data are shown for $t=L^2,1.5\,L^2, 2\,L^2$. Note that the distribution is proportional to $m^{-4/3}$ with $L$-dependent cutoff, and there is a condensate that detaches from the distribution and moves along the mass axis with time.} \label{ss-takayasu}
\end{center}
\end{figure}

With continuous injection of mass, at very large times $(t\gg L^2)$ the mass distribution consists of two parts: a time-independent distribution, $P(m)\sim m^{-4/3}$ with a system size dependent cutoff, coexisting with a condensate at a single site carrying a very large mass that grows linearly in time (Fig. \ref{ss-takayasu}). The cutoff mass for the `steady state' distribution is proportional to $L^3$, which fits in with the fact that in the coarsening regime the cutoff mass is $\sim t^{3/2}\propto \mathcal{L}(t)^3$, and the system reaches steady state when $\mathcal{L}(t)\approx L$. The existence of a steady state distribution coexisting with a condensate is reminiscent of the CMAM, with the difference that the condensate has a steady mean value $(\sim L)$ in the CMAM in contrast to the growing condensate in the Takayasu aggregation model.

From the distribution one can readily see that the average and the standard deviation of the total mass in the fluid are constants with values $\sim L^3$. The implications are: first, at such large times, the additional mass injected into the system accumulates in the condensate, and to the leading order the average mass of the condensate grows linearly in time. This may be understood as follows. In a single time unit a total mass $L$ is injected into the system, and therefore in a time $L^2$ the total amount of injected mass is $L^3$. Because of the local diffusive dynamics, large chunks of masses, including masses as large as $L^3$, can accumulate at locations other than the already grown condensate site in that time window. However, the condensate is also freely diffusing and would explore the system in about the same time. On encountering the newly grown condensate(s), it swallows it, thereby removing it from the fluid and rapidly growing its own mass. Thus the extra mass coming into the system is eventually taken up by the condensate. Further this implies that the standard deviation of the condensate mass as well as the total mass in the fluid are both proportional to $L^3$. These super-large values of the mean and standard deviation of the condensate were confirmed numerically (not shown here).

\section{\label{sec:level4} Dynamic condensates in the Takayasu aggregation model}

\subsection{Mass distribution during coarsening} %Indications from $P(m,t)$
In section \ref{sec:level2} we saw that in the CMAM models dynamic condensates play a major role in the evolution of the state during coarsening. Here, starting from an empty lattice, we investigate the evolution of the mass distribution in the Takayasu aggregation model. We take both $D$ and the rate of particle injection at each site to be $1$.

In Fig. \ref{Takayasu_configuration} the configurations at different times are shown, clearly suggesting the formation of dynamic condensates. At larger times there is a smaller number of larger condensates, separated from each other by roughly $\mathcal{L}(t)$.

\begin{figure}[H]
\begin{center}
    \includegraphics[width=7.5cm, height=5cm]{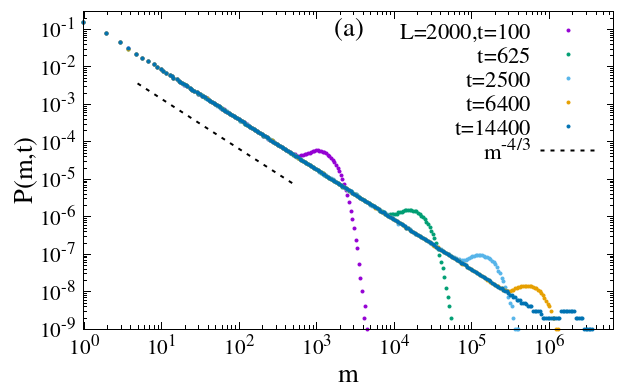} \hspace{1cm}
    \includegraphics[width=7.5cm, height=5cm]{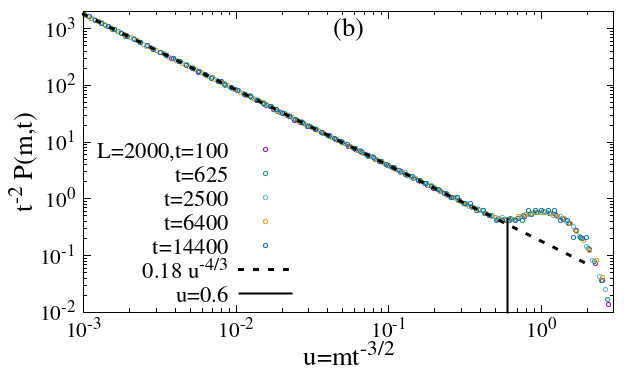}
\caption{\it \raggedright (a) The mass distribution in the coarsening regime of the Takayasu aggregation model. Note the bump at large masses, indicating the presence of dynamic condensates. (b) The collapsed distribution in the form of Eq. \eqref{scaling_takayasu}. The power law part is indicated by the dashed line. Note that, in the collapsed plot the bump starts approximately at $u=m\,t^{-3/2}\simeq 0.6$ (vertical line).}\label{pmt_takayasu}
\end{center}
\end{figure}

The mass distribution at large but finite times is shown in Fig. \ref{pmt_takayasu}. Fig. \ref{pmt_takayasu}(a) shows that $P(m)$ decays as a power law ($\sim 1/m^{4/3}$) up to a time dependent mass scale ($\sim t^{3/2}$), at which there is a prominent bump before the distribution falls to zero.
On defining $u=m/t^{3/2}$ and writing
\begin{equation}
 P(m,t)=t^{-2}\,f(u), \label{scaling_takayasu}
\end{equation}
we see that the curves in Fig. \ref{pmt_takayasu}(a) collapse onto a single curve (Fig. \ref{pmt_takayasu}(b)). Note however that the lower limit $u_0$ of $f(u)$ is time dependent, $u_0(t)\sim 1/t^{3/2}$ as the lowest value of mass is unity. Inspection of Fig. \ref{pmt_takayasu}(b) shows that, $f(u)\approx B\,u^{-4/3}$ with $B\simeq0.18$ in the range $u\in(u_0,u_c)$, where $u_c\simeq 0.6$. For $u>u_c$, $f(u)$ shows a bump, which decays exponentially as $u\rightarrow \infty$.

We associate the bump with dynamic condensates and ask for the fraction $x$ of the total mass they carry. It is simpler to evaluate the fraction of mass carried by the fluid. Recalling that the total mass in the system at time $t$ is $Lt$, it follows that
\begin{equation}
1-x \approx \, B\,\int_{0}^{u_c} u\,u^{-4/3}\,du \, \simeq 0.19.
\end{equation}
This implies that only about $20\%$ of the total mass is carried by the fluid while the condensates carry around $80\%$. This underscores the importance of the dynamic condensates in the Takayasu aggregation model.

\subsection{Condensates in regions of size $\mathcal{L}(t)$}

\begin{figure}[H]
\begin{center}
    \includegraphics[width=7.7cm, height=5.5cm]{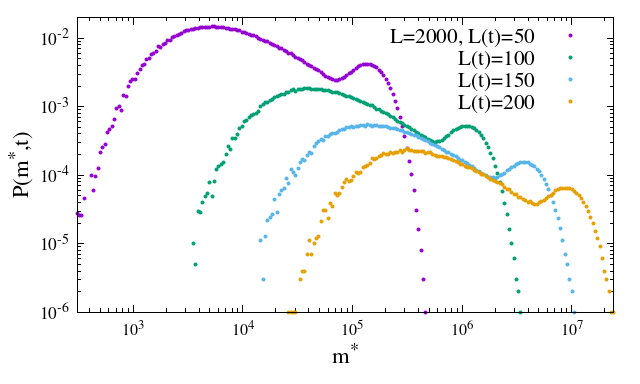} 
    \includegraphics[width=7.7cm, height=5.5cm]{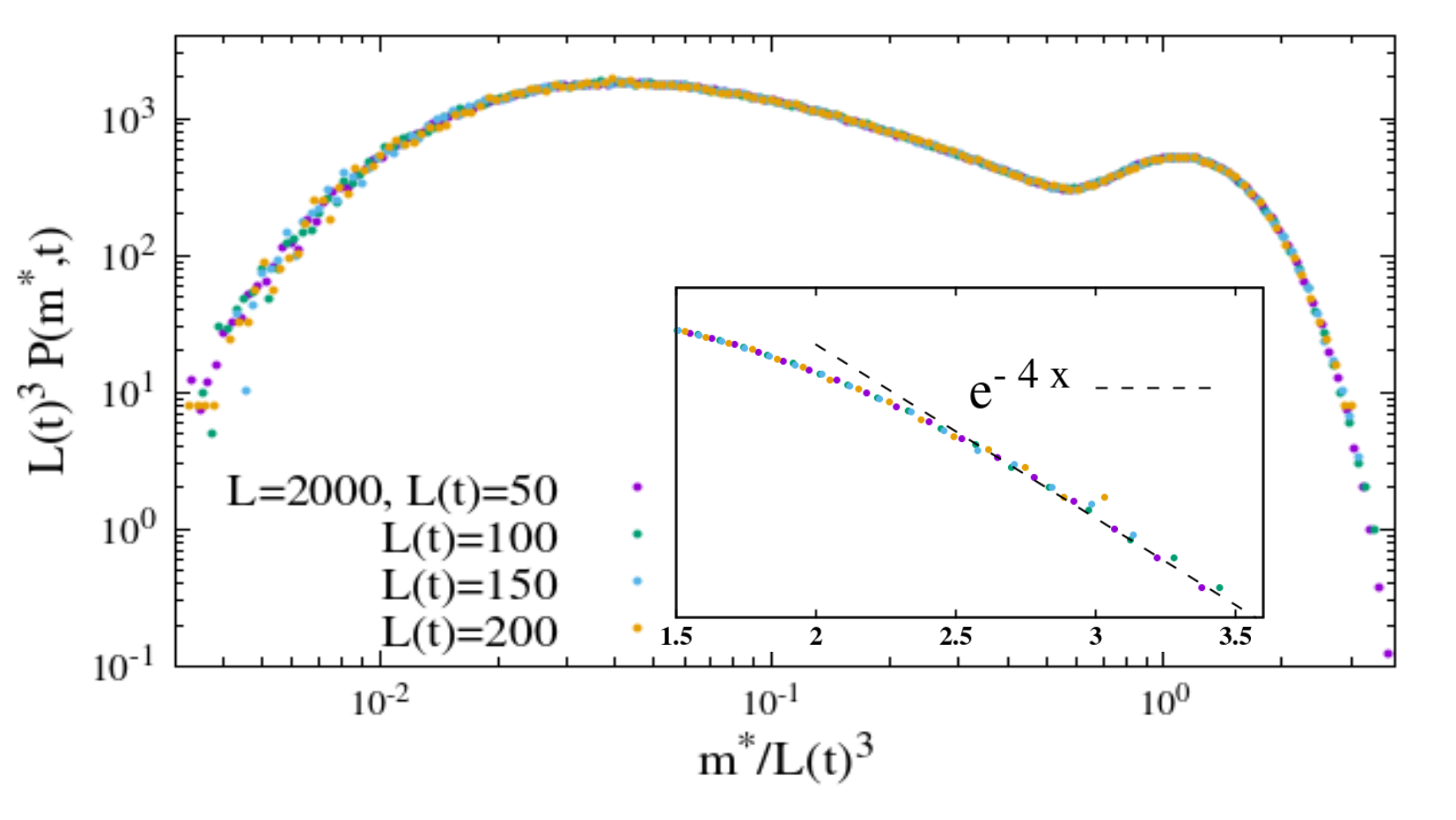}
\caption{\it \raggedright Left panel: The distribution of the largest subsystem mass in the growing regime of the Takayasu aggregation model. Right panel: The scaled distribution. Inset: exponential decay of the scaled distribution. Note the double peaked behaviour of the distribution which scales with $t^{3/2}\equiv \mathcal{L}(t)^3$.}\label{subsystem-max}
\end{center}
\end{figure}

The scaling of the correlation function in Fig. \ref{corr_takayasu} shows the existence of a growing length scale $\mathcal{L}(t)\sim t^{1/2}$, which is consistent with the structure of the configurations (Fig. \ref{Takayasu_configuration}), where the mean separation of the condensates is of the order $\mathcal{L}(t)$. 
To study the nature of the dynamic condensates, we divide the entire system of size $L$ into $N(t)=L/\mathcal{L}(t)$ subsystems, and numerically evaluate the behaviour of the largest mass $m^*$ in each subsystem. In the simulations we have taken $\mathcal{L}(t)=c\,t^{1/2}$ with $c=1$.
Interestingly, the entire distribution ${\rm Prob}(m^*,t)$ scales with $t^{3/2}\equiv \mathcal{L}(t)^3$: ${\rm Prob}(m^*,t)=t^{-3/2}\,F(m^*/t^{3/2})$, and the function $F(u)$ is bimodal (Fig. \ref{subsystem-max}). The above scaling form implies that the standard deviation as well as the other cumulants of the mass of the dynamic condensates are as large as the mean $(\sim \mathcal{L}(t)^{3/2})$, indicating a fluctuation-dominated coarsening state.

The bimodal nature of the distribution of the local largest mass signifies that as in CMAM, there are several segments of size $\mathcal{L}(t)$ in which there is no dynamic condensate. However, unlike the CMAM, here the entire mass distribution has the same scaling-like behaviour (Eq. \eqref{scaling_takayasu}), and consequently the full distribution of the local largest mass including those in the subsystems without dynamic condensates follows the same scaling. This is also consistent with the fact that the humps are separated at $u=m^*/t^{3/2}\approx 0.6$, the point at which the bump in the full mass distribution starts (see Fig. \ref{pmt_takayasu}).

\subsection{Multiplicative logarithms in the globally largest mass}
Now let us look into the properties of the largest mass $M^*$ in the entire system, closely following the argument used recently in a model of clustering driven by fluctuating surfaces \cite{cdmax}. In the Takayasu aggregation model, each of the $N(t)=L/\mathcal{L}(t)$ subsystems contributes roughly one dynamic condensate, and the globally largest mass is the largest of these. Since the subsystems are almost independent, so are the dynamic condensates. Therefore at any time $t$ such that $N(t)\gg 1$, the statistics of the globally largest mass reduces to the extreme value statistics of $N(t)$ independent random variables drawn from the distribution ${\rm Prob}(m^*,t)$. Now, from the numerical data we observed that the tail of the scaled distribution is exponential in nature (see Fig. \ref{subsystem-max}), implying that, $\lim_{m^*\rightarrow \infty} {\rm Prob} (m^*,t)\sim \exp(-m^*/ct^{3/2})$. Therefore the largest of the $N(t)$ samples drawn from ${\rm Prob}(m^*,t)$ follows a Gumbel distribution \cite{Tippett,Gumbel} with mean,
\begin{equation}
 \langle M^* \rangle \approx t^{3/2}[a+b\,\ln(N(t))] = t^{3/2}[a+b\,\ln(L/t^{1/2})]. \label{mmax}
\end{equation}

\begin{figure}[H]
\begin{center}
    \includegraphics[width=12cm, height=8cm]{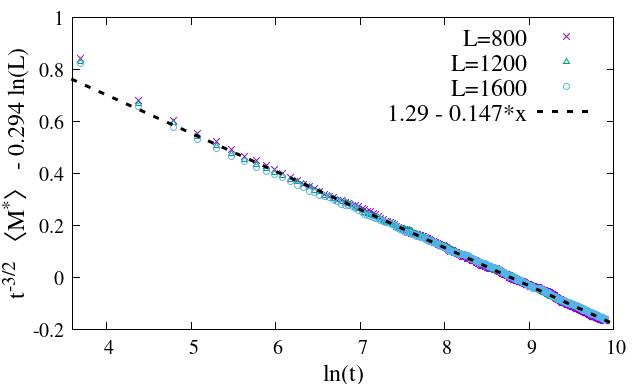}
\caption{\it \raggedright The mean of the globally largest mass for different system sizes as a function of time. The logarithmic behaviour of $\langle M^* \rangle$ is consistent with Eq. \eqref{mmax}, with $a=1.29$ and $b=0.294$.}\label{global-max}
\end{center}
\end{figure}

Equation \eqref{mmax} shows that the mean largest mass has a strong logarithmic dependence on both time and system size.
To check the expression numerically, we proceed as follows. Note that, $\langle M^* \rangle/t^{3/2}$, when plotted against $\ln(t)$, is a straight line with slope $b/2$ at large times. Once the slope and therefore $b$ is thus determined from data, then $\langle M^* \rangle/t^{3/2} - b\,\ln(L)$ can be plotted for for different values of the system size $L$, which should be independent of $L$ and collapse on each other.
This is indeed corroborated by simulation results, as shown in Fig. \ref{global-max}.

\section{\label{sec:level5} Conclusion}
Historically, the Takayasu aggregation model has been the source of a rich variety of ideas and connections in nonequilibrium statistical physics. In this paper we have discussed some of the earlier results and connections, and also some new results in one dimension. These are summarized below.

\paragraph{I.}
The first result concerns the distribution of masses at time $t$. It is well known that the distribution of masses in the Takayasu aggregation model follows a $m^{-4/3}$ law in the limit $t \rightarrow \infty$ in an infinite system, but it is not well known that at any finite but large time, the distribution also includes a prominent maximum which signifies the existence of dynamic condensates. We confirmed this through a Monte Carlo study by monitoring the largest mass clusters in segments of length $\mathcal{L}(t) \sim t^{1/2}$, a technique which was employed fruitfully earlier to study condensates in conserved mass aggregation models \cite{Iyer}. The result for the Takayasu aggregation model is surprising --- a very large fraction of the mass is held in dynamic condensates, and much less in the power law part. Further, the fact that the distribution of the dynamic condensates follows a simple scaling form implies large fluctuations, with the standard deviation of dynamic condensate mass being proportional to the mean $\sim t^{3/2}$.

\paragraph{II.}
The second result pertains to the global maximum of the mass in a large but finite system of length $L \gg {\cal L}(t)$.
This is obtained by applying extremal statistics to the set of dynamic condensates which are essentially uncorrelated. The crucial point is that the number of condensates depends on the time and system size. This induces multiplicative logarithms involving the size and time which are well verified by our Monte Carlo results.

\paragraph{III.} At large times ($\mathcal{L}(t)\sim L$ and larger), the system reaches a steady state exhibiting coexistence of power-law distributed critical fluid and a giant condensate. The extra mass injected into the system is eventually swallowed by the condensate resulting in a linear growth of the condensate mass and a massive fluctuation of $\sim L^3$. This is the third main result.

In conclusion, we recall that the Takayasu aggregation model has correspondences with several other models and processes in statistical physics. The implications of our results for some of these problems will be taken up elsewhere. It would be interesting to study the behaviour of extremal quantities in the generalised models discussed in section \ref{sec:level3}, namely the Takayasu process with opposite charges, the in-out model at its critical point, and some other models \cite{Raissa-dragon-23}.

\paragraph*{Acknowledgements:} We acknowledge R. Rajesh for several useful inputs and Chandrashekar Iyer for discussions. M.B. acknowledges support under the DAE Homi Bhabha Chair Professorship of the Department of Atomic Energy, India. This project was funded by intramural funds at TIFR Hyderabad from the Department of Atomic Energy (DAE), India.

\end{document}